\documentclass[10pt,preprint]{aastex}
\begin{document}
\slugcomment{To be submitted to The Astronomical Journal}
\title{The ACS Survey of Galactic Globular Clusters. X. New Determinations of Centers for 65 Clusters.}
 
\author{Ryan Goldsbury\\
\affil{Department of Physics and Astronomy, University of British Columbia, Vancouver, BC, Canada\\
Email:{\tt rgoldsb@astro.ubc.ca}}}

\author{Harvey B. Richer\\
\affil{Department of Physics and Astronomy, University of British Columbia, Vancouver, BC, Canada\\
Email:{\tt richer@astro.ubc.ca}}}

\author{Jay Anderson\\
\affil{Space Telescope Science Institute, Baltimore, MD, USA\\
Email:{\tt jayander@stsci.edu}}}

\author{Aaron Dotter\\
\affil{Space Telescope Science Institute, Baltimore, MD, USA\\
Email:{\tt dotter@stsci.ca}}}

\author{Ata Sarajedini\\
\affil{Department of Astronomy, University of Florida, Gainsville, FL, USA\\
Email:{\tt ata@astro.ufl.edu}}}

\author{Kristin Woodley\\
\affil{Department of Physics and Astronomy, University of British Columbia, Vancouver, BC, Canada\\
Email:{\tt kwoodley@phas.ubc.ca}}}

\shortauthors{Goldsbury {\it et al.}}
\righthead{Globular Cluster Centers}

\begin{abstract}

We present new measurements of the centers for 65 Milky Way globular clusters.  Centers were determined by fitting ellipses to the density distribution within the inner $2\arcmin$ of the cluster center, and averaging the centers of these ellipses.  The symmetry of clusters was also analyzed by comparing cumulative radial distributions on opposite sides of the cluster across a grid of trial centers.  All of the determinations were done with stellar positions derived from a combination of two single-orbit ACS images of the core of the cluster in $F606W$ and $F814W$.  We find that the ellipse-fitting method provides remarkable accuracy over a wide range of core sizes and density distributions, while the symmetry method is difficult to use on clusters with very large cores, or low density.  The symmetry method requires a larger field, or a very sharply peaked density distribution.       

\end{abstract}


\section{Introduction}

\noindent
The need for precise centers of Galactic globular star clusters is more critical today than ever before.  For example, there is much work being done searching for intermediate-mass black holes (IMBHs) in globular cluster cores.  The masses of these IMBHs are expected to be quite small compared to their galactic counterparts, a few hundred to a few thousand $M_{\odot}$, and so the distance from the center of the black hole within which the stars would be observably influenced is also quite small.  As shown in Anderson \& van der Marel (2009), a center that is off by only $12\arcsec$ can greatly affect conclusions about the presence of an IMBH.  Additionally, recent work analyzing the radial distributions of various stellar populations within globular clusters, such as Bellini et al. (2009) and Ferraro et al. (2009), also depends on the precision of the centers used.  An incorrect center will dilute the differences between any two radial distributions centered on the same erroneous location.

\noindent
Historically, coordinates for the centers of globular clusters have been determined by a variety methods, and have been compiled in a catalog made available online by Harris (1996).  The parameters in this catalog are still widely used today.  The majority of the centers in this catalog come from a list compiled by Djorgovski and Meylan (1993).  Of the 143 globular clusters contained in this 1993 list, 109 of the center coordinates come from Shawl and White (1986).  The centers from Shawl and White were taken from scanned photographic plates.  These scans were smoothed and the center was found with an algorithm known as the SWIRL technique, developed by C.R. Lynds (described in Shawl and White 1980).  This method involves analyzing the density over a number of regions symmetrically selected in the X and Y directions, and manually adjusting the center point until the densities in the corresponding X and Y regions are as similar as possible.  Uncertainties were given as the standard error of repeated determinations with this method.  These center coordinates have stood the test of time quite well, but with the advancement of modern CCDs and better reduction techniques, more precise centers can be determined.  

\noindent
Whereas previous determinations were made on photographic plates and focused on the outskirts of clusters, our measurements here will focus on the central regions.  If there are any asymmetries in the stellar distribution, the two methods will not arrive at the same center.  The SWIRL method analyzes the symmetry of light from the entire cluster, but has no way to deal with asymmetries in the distribution caused by small numbers of giant stars.  We caution that the method discussed here can only be used to determine the center of the starcount distribution.  The center of mass must be determined through analysis of the dynamics of the cluster.     

\section{Data Reduction}

\noindent
All of the data used are from the ACS Survey of Galactic Globular Clusters (Sarajedini et al. 2007).  A thorough discussion of the reduction methods used for the data can be found in Anderson et al. (2008).  A brief summary follows.

\noindent
Each cluster was observed with HST's ACS/WFC for one orbit through each of $F606W$ and $F814W$.  Each orbit contained one short exposure (to fill in the saturated stars) and four to five deep exposures, which were stepped over the ACS chip gap for more uniform spatial coverage.  The catalog was constructed by analyzing the field patch by patch in all the individual exposures simultaneously.  In each patch, the brightest stars were found first and fitted with a PSF that was tailored to each particular exposure, then subtracted to allow fainter neighbors to be found.  Only stars that clearly stood out above the known subtraction errors and known PSF artifacts in multiple exposures in both filters were included in the catalog.  The very few stars that were saturated even in the short exposures were measured by fitting the PSF to the surrounding unsaturated pixels.  The 2MASS catalog was to used convert the positions measured in the image-based frame into absolute RA and Dec.  The operations that follow in this paper were based on the Right Ascension and Declination coordinates, as well as the $F814W$ magnitude for each star.  This entire data catalog will soon be made public (Sarajedini et al. 2011).

\section{Methods for Determining The Center}

\noindent
Two methods for determining the center of the starcount distribution were employed.  Both methods begin with the positions and magnitudes from the catalog, constructed as described in Section 2.  The density method searches for the average center of a number of isodensity contours by fitting ellipses to these contours.  The pie-slice method searches for the most symmetric point of the cluster by analyzing the distributions of stars as a function of radius on opposite sides of the cluster.

\noindent
All of the centers reported in Table 1 were derived with the density contour method, as it was found to be more reliable over a wide range of density distributions than the pie-slice method.  Both of these methods are thoroughly discussed in Anderson \& van der Marel 2009, but they will be summarized here as well.  
 
\subsection{Density Contours}

\noindent
We began with the catalog as constructed above and converted the apparent $F814W$ magnitudes into an absolute magnitude scale using the distance moduli for each cluster from the Harris Catalog (1996).  All stars with an absolute $F814W$ magnitude fainter than 8 were immediately excluded.  This has the effect of smoothing out the distribution, as stars fainter than this are not likely to be found near much brighter stars, and they are very incomplete in the cores of denser clusters.  When stars of all magnitudes are included in the density distribution, many underdense regions around bright stars appear.  These holes do not reflect the true underlying density distribution, and so an arbitrary cutoff was implemented to limit these effects (Figure 1).  For denser clusters, stricter magnitude cutoffs were used.  The densest clusters required that only stars brighter than an absolute $F814W$ magnitude 2 be used, in order to remove the appearance of these holes.  These regions of incompleteness begin to significantly overlap as you approach the center of the cluster.  The number density of very bright stars in the cores of clusters is so large, that very faint stars are almost impossible to reliably detect.  As a result, fainter stars are almost entirely incomplete in the inner region of the cluster.  Because density is a function of radius, and the ability to reliably detect a star is dependent on the density of the bright stars in the region surrounding that star, the incompleteness of fainter stars is also a function of radius.  So, although incompleteness is also present to a lesser extent in stars above the implemented cutoff, the radial dependence of the incompleteness means that there will be no directional bias introduced by the incompleteness, as evidenced in Figure 2, and center determinations should not be affected.   

\noindent
The coordinate system used for the determination is oriented to Right Ascension and Declination, but with the necessary cosine term to generate a projection onto the plane tangent to the sky at the center of the field.  The density of the cluster was then constructed over a grid of points.  This grid was centered on the RA and Dec. from the Harris Catalog (1996). The grid extends $100\arcsec$ in each direction, sampling every $2\arcsec$.  As mentioned in Anderson \& van der Marel 2009, significant overbinning is required to generate a smooth distribution.  Circular bins were used with a radius of $25\arcsec$ in most cases.  However, clusters with very flat distributions required larger bins (up to a radius of $40\arcsec$) to smooth the distribution.  To remove the effects of the edge of the field on the density distribution, the density values for points on the sample grid that were within a bin radius of the edge of the field were set to 0, so as not to be included in any of the contours.  That is to say, if the circular bin at that point included area outside of the field, that point was excluded from the ellipse fitting portion of this method.   

\noindent
After the distribution was generated and the edge effects were accounted for, the distribution was broken up into eight contours.  The contours were spaced evenly between the minimum non-zero density and the maximum density.  The outer three contours were not used due to the fact that they were often quite azimuthally incomplete, and this could potentially bias the ellipse fits.  The innermost contour was also ignored as it is actually a solid two dimensional region of points and could not be fit well with an ellipse.  Each of the four remaining contours was then fit with an ellipse.  This step is outlined in Figure 3.  For most of the denser clusters, it is possible to fit many more than four ellipses (in some cases up to 20).  However, larger numbers of smooth contours cannot be generated for the clusters with low numbers and/or flat distributions.  In order to maintain consistency, eight contours and four ellipses were adopted as the maximum numbers that could be used for every cluster. The final center value was determined as the average center of these four ellipses. The uncertainty was estimated as the standard deviation of the ellipse centers.  A plot of the density contours as well as the ellipses fit to these contours and their centers for the cluster NGC 1261 is shown in Figure 4.  Plots such as this one are available for all 65 clusters in Table 1 as part of the supplementary material.

\noindent
The consistency of the method discussed above as well as the error estimates for these measurements were analyzed through bootstrapping.  Synthetic cluster distributions were generated and analyzed with the density contours method.  The number of stars in the synthetic distributions ranged from 25,000 to 250,000 in increments of 25,000.  These values cover roughly the range of the 65 clusters presented in this paper after the initial faint magnitude exclusion.  For each increment between 25,000 and 250,000, one hundred distributions were generated, and their centers were measured through the density contours method.  The standard deviation of the centers used in each determination, as well as the actual errors from the true center of the generated distributions were averaged over these 100 samples.  The estimated uncertainty and actual error are plotted against the number of stars in Figure 5.  This figure suggests that the standard deviation of the ellipse centers provides a very good estimate of the actual uncertainty in the determination.

\noindent 
A distribution of the differences in X and Y from the catalog centers is shown in Figure 6.  X and Y are oriented to RA and Dec., and represent seconds of arc on the sky.  The center coordinates were also compared to those presented in Noyola \& Gebhardt (2006).  There appears to be no systematic difference between the centers determined in this paper and those determined by Noyola \& Gebhardt.  The means of the differences in RA and Dec. are consistent within the standard errors with a distribution about the origin.  With respect to the centers in the Harris Catalog, the distribution of differences in X is consistent with a random normal distribution centered around zero, suggesting that there are no systematics biasing the centers in RA.  However, the average difference in the Y direction is greater than the standard error of the differences in that direction, indicating that the distribution is just barely inconsistent with a center of zero.  This suggests a possible systematic difference of about $1.7\arcsec$ in the Dec. direction between the centers provided in Table 1, and the centers given in the Harris Catalog (1996).  The centers presented by Noyola \& Gebhardt also exhibit this systematic difference with a distribution of differences from the Harris Catalog centered approximately $1\arcsec$ South of the origin.

\subsection{Pie-Slice Contours}

\noindent
To begin, a grid of sample points was constructed around the center value from the Harris Catalog.  The grid extends $40\arcsec$ in each direction, sampling every $2\arcsec$.  At each point on the grid, the stars within a radius of $1.2\arcmin$ were divided into eight different pie slices corresponding to the cardinal and semi-cardinal directions in RA and Dec. (see Figure 7).  A cumulative radial  distribution was then generated for each of the eight pie-slices.  The four pairs of opposing distributions were then compared.  Figure 8 shows two pie-slices and the corresponding cumulative distributions of the stars within these regions.  To compare the distributions at a single point, the integrated area between the curves was used.  Four pairs of distributions were compared, giving four values for each point on the grid.  These values were then added together to yield a measure of the symmetry of the cluster at that point on the sample grid.  When determined in this way,  higher values correspond to less symmetry, as this indicates a larger difference in the opposing distributions at that point.  This was done for every point on the sample grid.  The result is a two-dimensional array of values that describe the symmetry of the cluster as a function of position in projected space.

\noindent
As discussed in the previous section and shown in Figure 1, bright giant stars create large holes in the apparent stellar distribution.  These holes are often asymmetrically distributed due to their small numbers, and as a result large differences appear in opposing radial distributions at locations that are otherwise quite symmetric.  Because of this effect, many of the denser clusters required that an incompleteness mask be applied before using the pie-slice method.  In this case, simply imposing a strict magnitude cutoff does not work, as this lowers the number of stars in each pie slice to the point where the sample sizes are no longer useful. This is discussed in more detail in Anderson \& van der Marel 2009 (see Fig. 12 of that paper).

\noindent
It is important to note that a minimum of two pairs of orthogonal pie slices must be used to get a center determination with this method, as each pair of pie slices is only sensitive to changes along one axis.  This is demonstrated in Figure 8.  Moving a pair of pie slices perpendicular to their orientation keeps them in a position that is still symmetric along their axis.  For this reason, a second pair of pie slices must be used to sample the symmetry in the direction orthogonal to the first.  For our sample of clusters, it was found that eight pie slices worked better than four, however switching to sixteen pie slices resulted in sample sizes that were too small to be useful.

\noindent
Determining the size of pie slices to use also takes some consideration.  The radius of the pie slices must be determined based on the size of the field, and the size of the sample grid.  It is important to constrain the radius so that no point on the grid is affected by the edge of the field.  It is also important that the radii of the pie slices remain constant across the grid, so as not to bias the determination.  This creates a problem, as larger radii create larger samples and thus better distributions, however the sample grid must then be smaller so that the pie slices at the outer edges of the grid do not run out of the field.  For this same reason, sampling over a larger area requires the radii of the pie slices to be smaller.  

\noindent
After the symmetry value is calculated at each point in the sample grid, contours can be created and a center determined by finding the most symmetric point on the grid.  Due to the orientations of the pie slices, these contours will be eight sided figures (or $2N$-sided figures for $N$ pairs of pie slices), and so it is not appropriate to fit them with ellipses as discussed in the density contours method.  This method can only show the most symmetric point on the sample grid, and so the uncertainty of this center is limited by the spacing of the grid points.  Even for centrally concentrated clusters with good numbers across the field, the symmetry did not appear to change significantly on scales smaller than $2\arcsec$, and so a grid spacing of less than this was not used.  This makes the centers determined by this method considerably less precise than those determined by the density contours method.

\section{Conclusions}

\noindent
In this paper we have presented new centers for 65 Milky Way globular clusters and outlined the methods used to determine them.  All of the centers in are given in Equitorial and Galactic coordinates in Table 1.  These centers are significantly more precise than what is currently available in the literature with an average uncertainty of less than $1\arcsec$.  In addition, many of these centers differ significantly from the values in Harris 1996.  Of the 65 clusters analyzed, 26 differ by more than $5\arcsec$, and 8 clusters differ by more than $10\arcsec$.  Precise centers are key to analyzing the dynamics in the central regions of clusters, and in particular, searching for evidence of IMBHs and analyzing the distributions of separate stellar populations within a cluster.  We have described the general outline of the density contours method and shown that incompleteness in faint stars should not bias this determination due to the dependence of incompleteness on the distance from the cluster center.  We have also discussed a separate method that relies on analyzing the symmetry of the clusters to find the center, however the dependence of this method on sample size and cluster size relative to the imaging field make it difficult to apply in most cases.  Denser clusters that require masking bright stars can make the symmetry method even more time consuming, and so we suggest that the density contours method is the preferred method of determing globular cluster centers.  

\acknowledgements 

This work was supported by the Natural Sciences and Engineering Research Council of Canada (H.B.R., A.D.), by NASA through HST grant GO-10775 (J.A., A.S.), and by CITA (A.D.). This research is based on NASA/ESA Hubble Space Telescope observations obtained at the Space Telescope Science Institute, which is operated by the Association of Universities for Research in Astronomy Inc. under NASA contract NAS5-26555.  This paper uses data from the ACS Survey of Galactic Globular Clusters (GO-10775; PI: Ata Sarajedini).

$Facilities$: HST (ACS)
 
\clearpage
\begin{figure}
\includegraphics{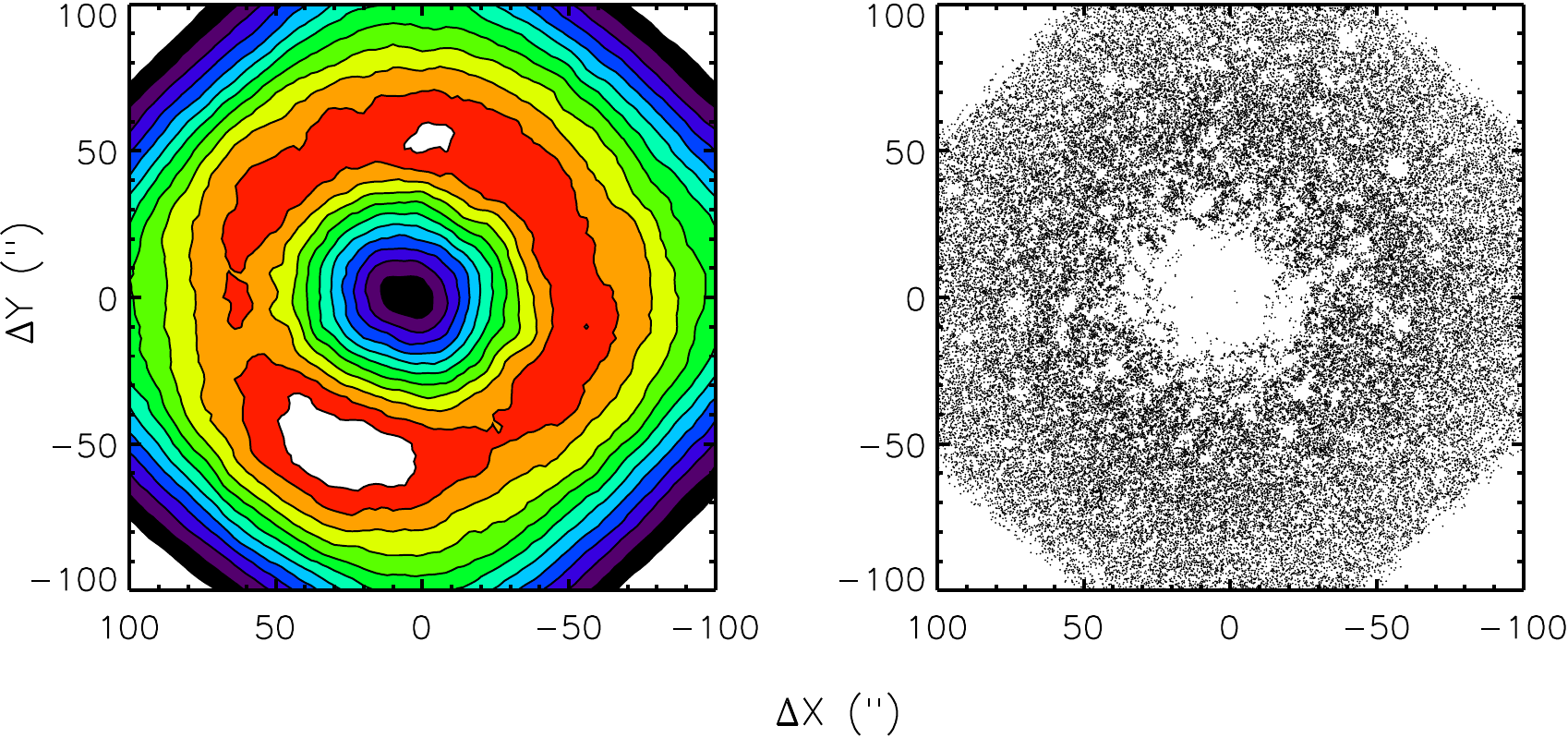}
\caption{Cluster NGC 1851. The density contours (left) of the stars that were rejected from the center determination method due to the imposed limiting $F814W$ magnitude cutoff, and positions of the stars in the field (right) after the initial data reduction (star positions are in black).  Large voids are seen near brighter stars, as these make it difficult to find fainter stars nearby.  The center is almost entirely empty, as it is far too dense to reliably find stars fainter than absolute magnitude 8.  The coordinates are in projected distance with respect to the center in the Harris Catalog (1996).}
\label{fig1}
\end{figure}

\clearpage
\begin{figure}
\includegraphics{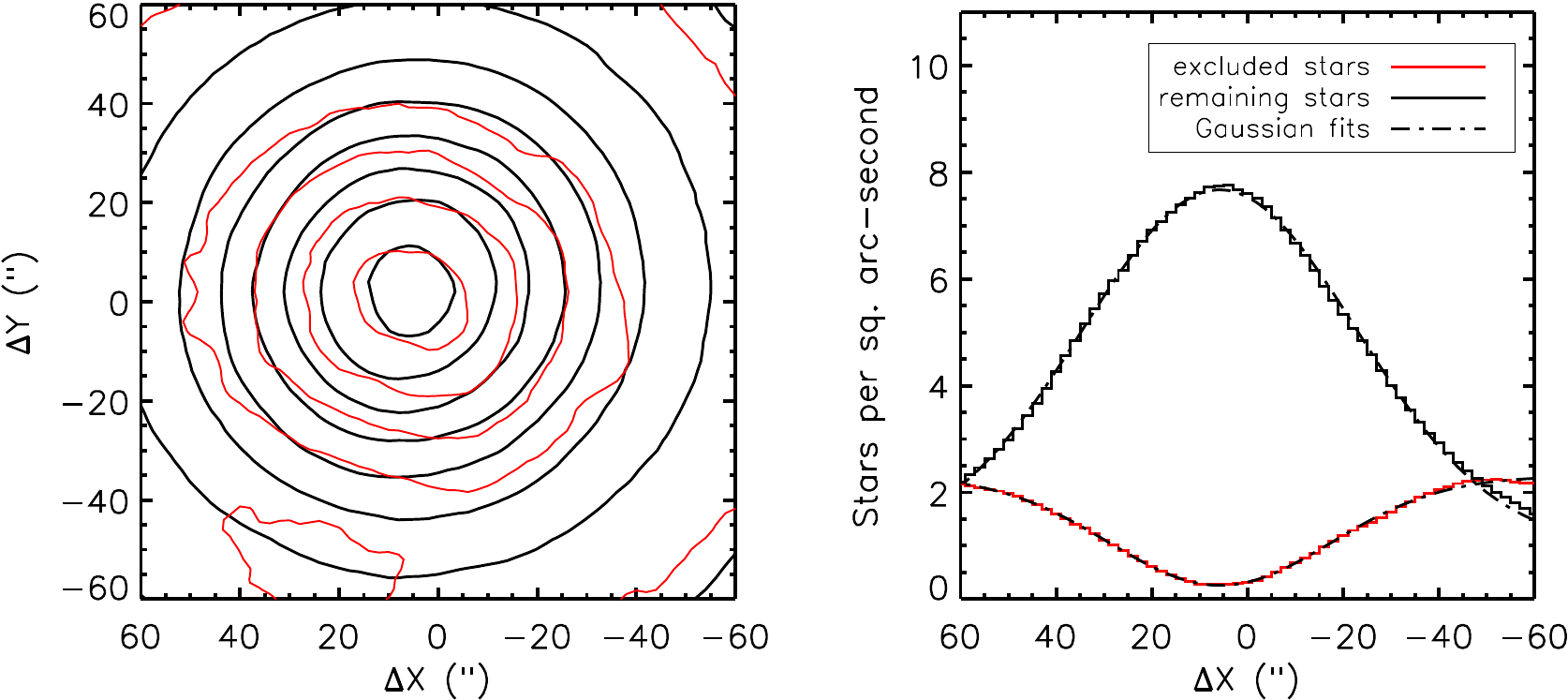}
\caption{On the left, density contours for NGC 1851 from the method outlined in section 3.1 are shown in black (thick).  Overplotted in red (thin) are the inner contours from the valley shown in Figure 1.  Ellipses fit to each of these sets of contours find average centers that agree to within $1.6\arcsec$ (within the estimated uncertainty of the two centers).  On the right, slices of the density surface are shown through $\Delta$$Y=0$ for the faint stars excluded, and the remaining stars used in the center determination.  Gaussians fit to these distributions find centers that agree to within $0.12\arcsec$.  This agreement supports the conclusion that, due to the radial dependence of incompleteness, this effect will not bias the center determinations.  The coordinates are in projected distance with respect to the center in the Harris Catalog (1996).}
\label{fig2}
\end{figure}

\clearpage
\begin{figure}
\includegraphics{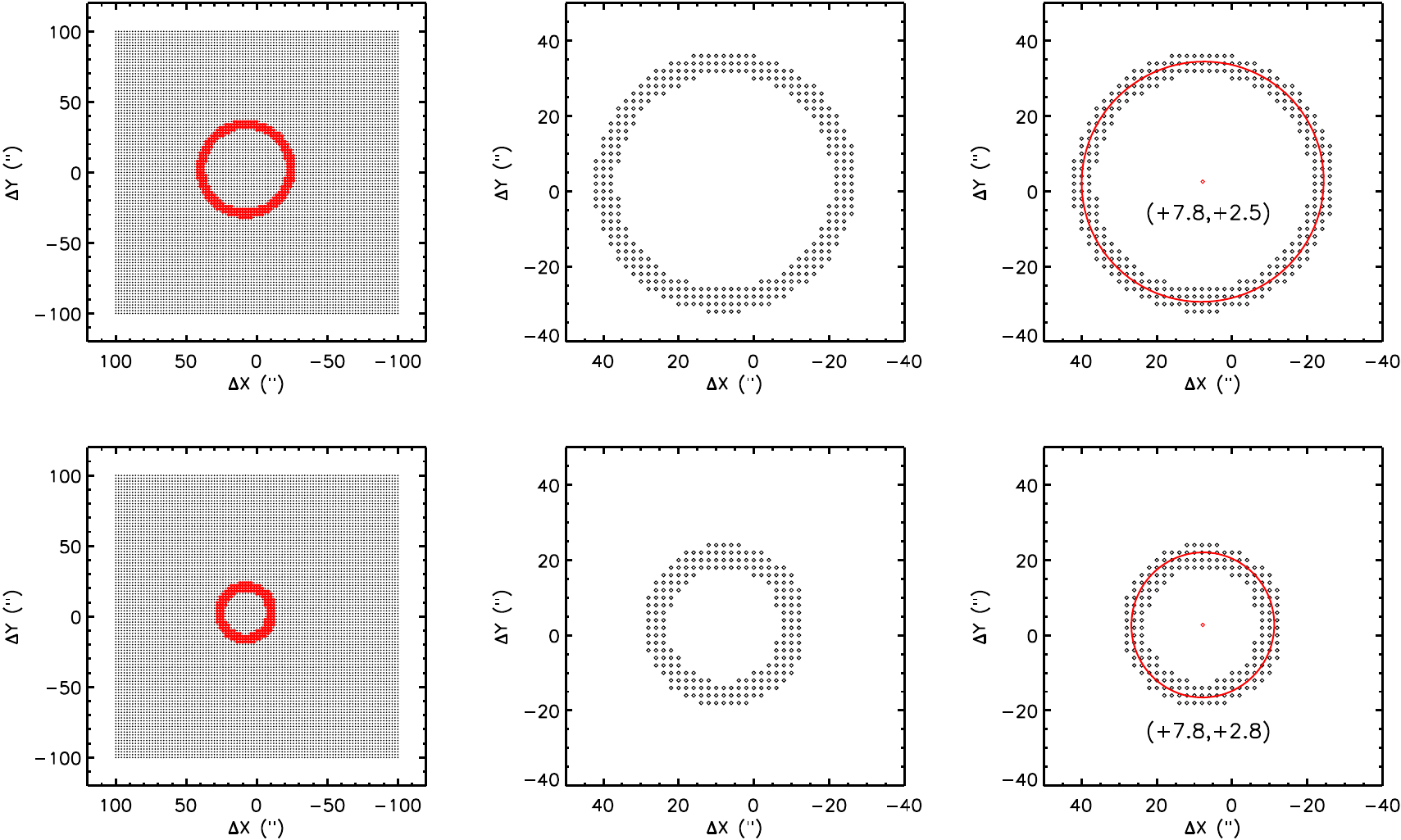}
\caption{Cluster NGC 1851.  A visual representation of the method described in section 3.1.  Each row shows the points on the sample grid that fall within a specified density bin, as well as the ellipse fit to these points, and the center for this ellipse.  The top row corresponds to a density bin of $\sim$2.5-3.3 stars per square arcsecond, while the bottom row corresponds to a bin of $\sim$4.0-4.8 stars per square arcsecond.  The coordinates are in projected distance with respect to the center in the catalog (Harris 1996).}
\label{fig3}
\end{figure}

\clearpage
\begin{figure}
\includegraphics{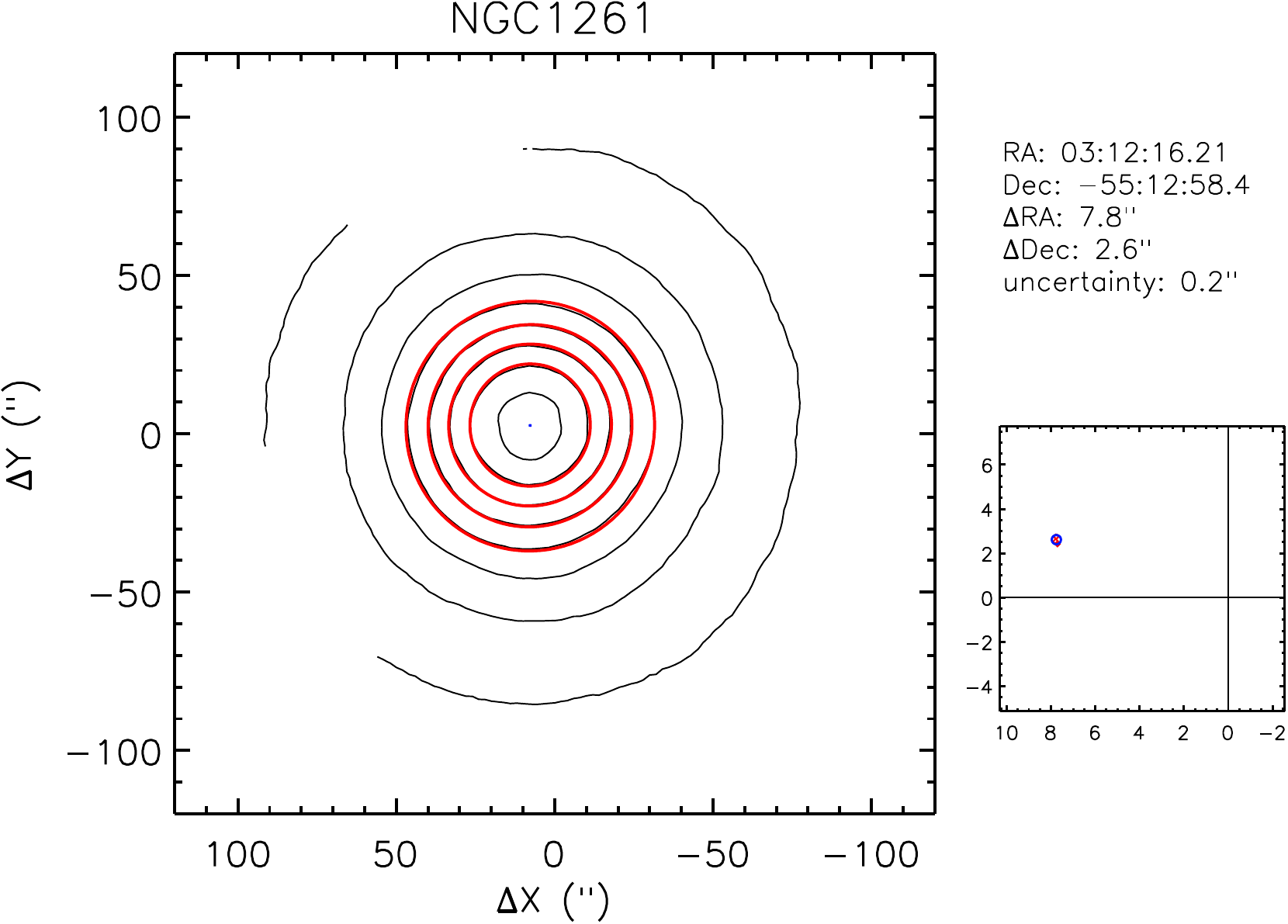}
\caption{Cluster NGC 1261.  Density contours are shown in black (thin).  Fit ellipses are shown in red (thick) with the centers as red dots.  The calculated center is shown as a blue circle with the radius equal to the estimated uncertainty.  Distances are given in seconds of arc in projected distance from the catalog center.  Relevant information is contained in the table to the upper right including the center in Right Ascension and Declination, the difference from the catalog center in seconds of arc in the RA and Dec. directions, and the uncertainty of the center determined as the standard deviation from all of the ellipse centers.  The smaller plot in the lower right shows a close-up of the determined center as well as the Harris center, which is [0,0] in this coordinate system.}
\label{fig4}
\end{figure}

\clearpage
\begin{figure}
\includegraphics{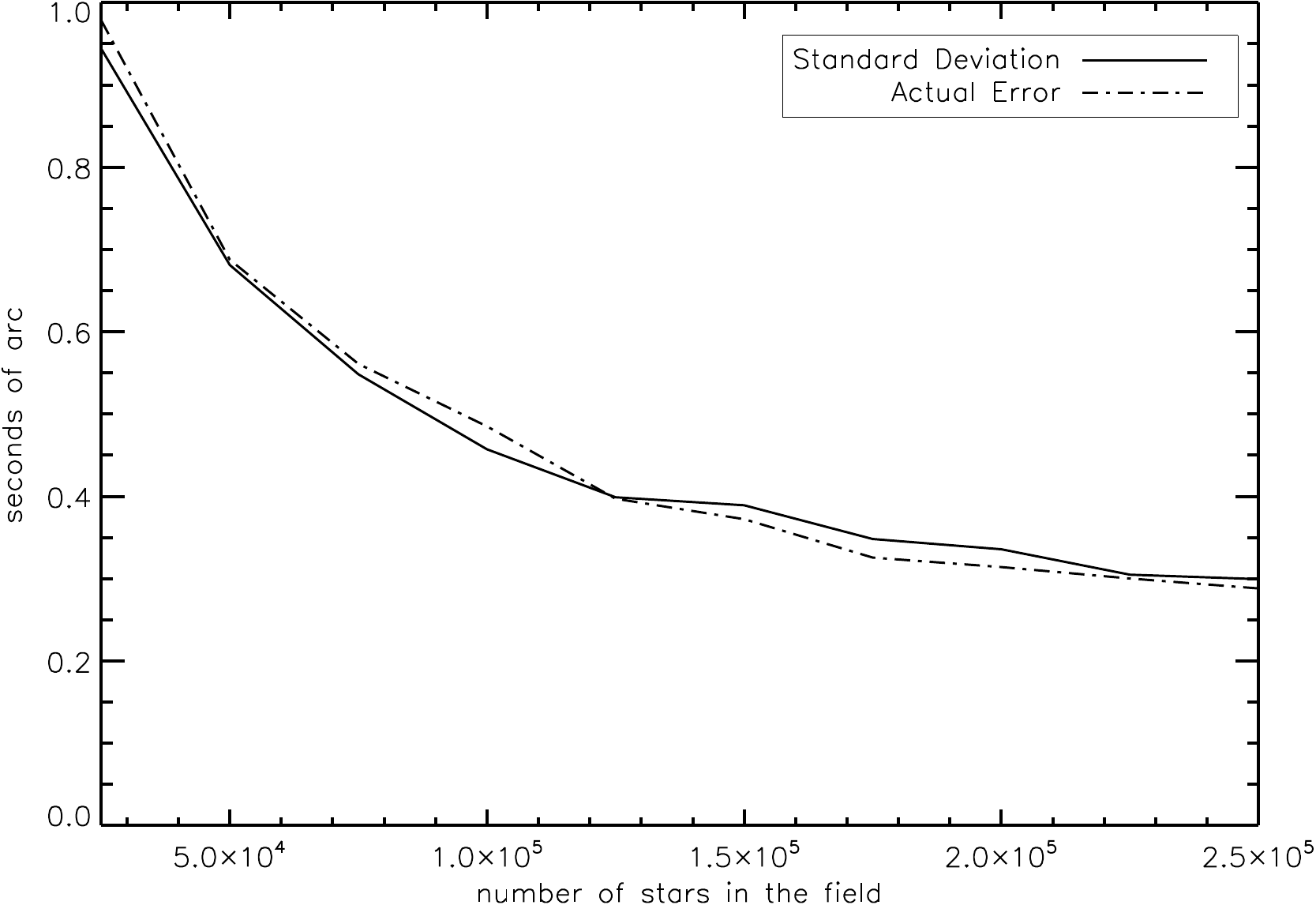}
\caption{The standard deviation (used as the estimated uncertainty) and actual error as a function of the number of stars in the field.  Values were determined through bootstrapping.  The standard deviation of the ellipse centers appears to be a good estimate for the uncertainty in the center.}
\label{fig5}
\end{figure}

\clearpage
\begin{figure}
\includegraphics{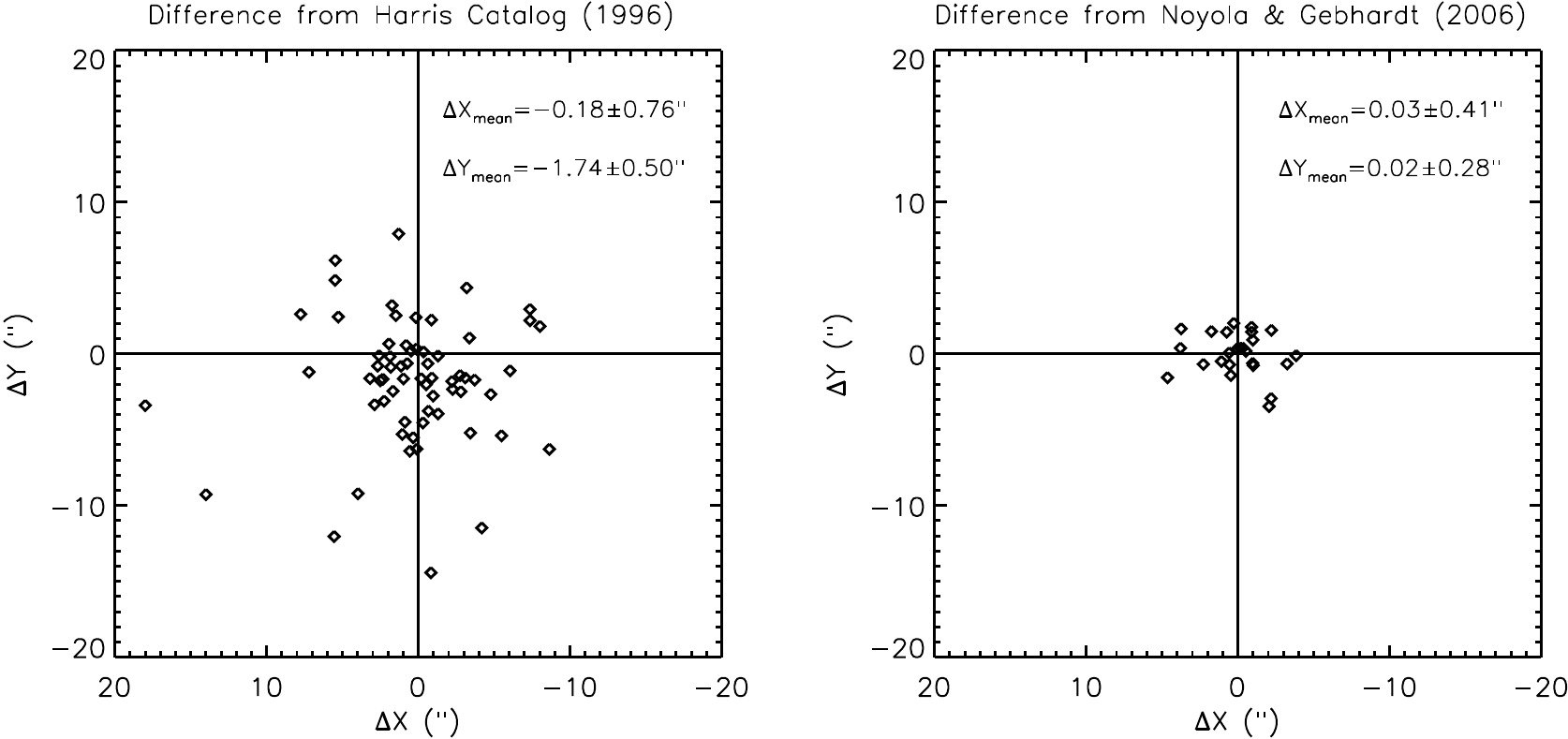}
\caption{The scatter of the differences between the determined centers and the center values given in the Harris Catalog (1996) and Noyola \& Gebhardt (2006).  The X and Y axes correspond to seconds of arc in the RA and Dec. directions.  The clusters NGC 7089 and NGC 6496 are not included in the left plot as they differ by greater than 20$\arcsec$, however they are still included in the average and standard deviation.  Of the 65 centers presented in this paper, 24 were also determined by Noyola \& Gebhardt (2006).  The differences between these 24 clusters are shown on the right.  The uncertainties given in each case are the standard error of the mean.  With respect to the Harris Catalog, the standard deviation of the scatter in RA is $6.65\arcsec$ and the standard deviation of the scatter in Dec. is $7.03\arcsec$.}
\label{fig6}
\end{figure}

\clearpage
\begin{figure}
\includegraphics{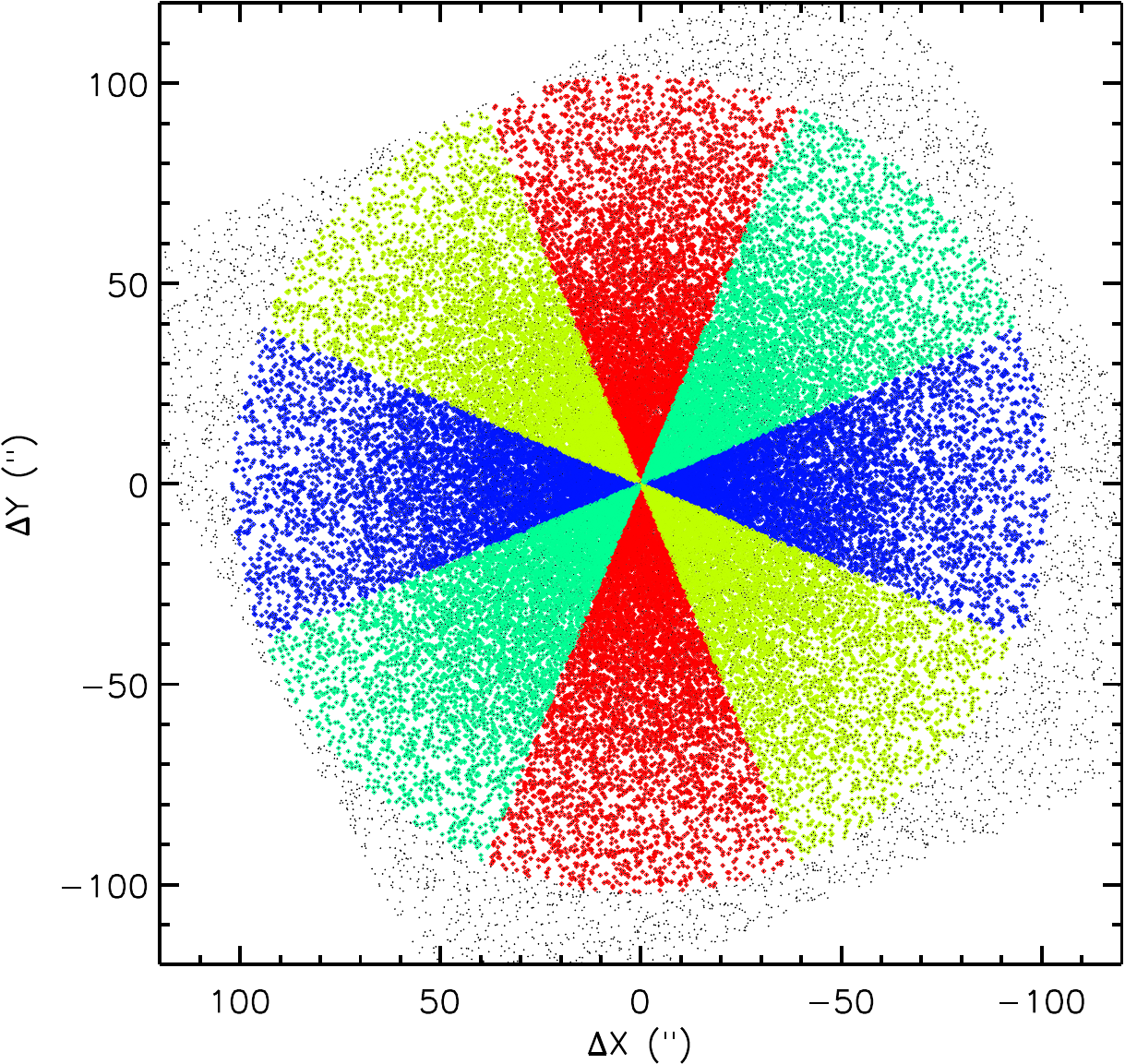}
\caption{The orientation of the eight pie slices used in the symmetry method are shown here.  Each pair is color coded.  The cumulative distributions of like colors are compared at each point on a sample grid.  The coordinates are in projected distance with respect to the center in the catalog (Harris 1996).}
\label{fig7}
\end{figure}

\clearpage
\begin{figure}
\includegraphics{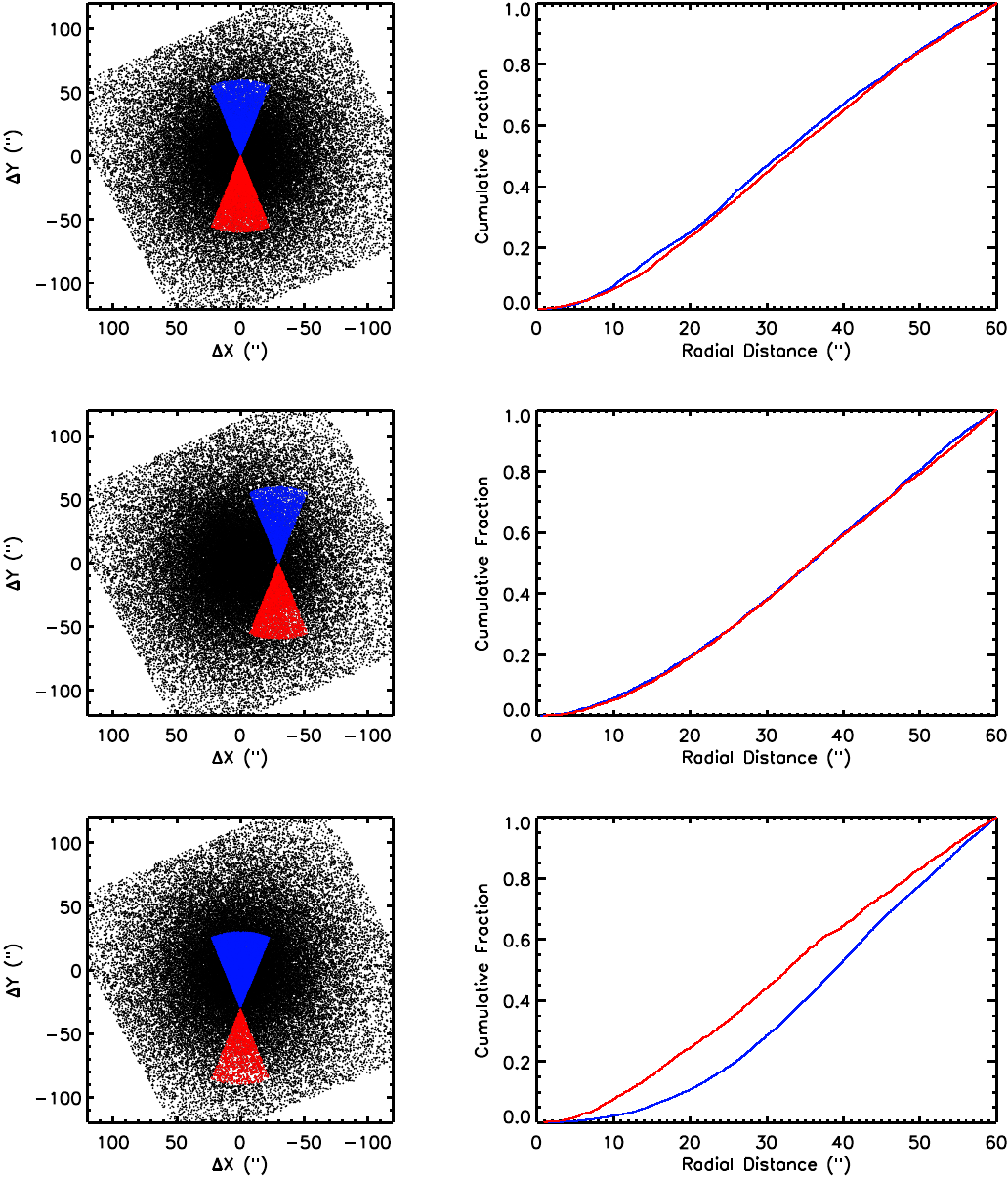}
\caption{Cluster NGC 1261.  The North and South pie slices as well as their cumulative distributions are shown at three different locations in the cluster.  The top row corresponds to the center found by the density contours method.  The middle row corresponds to a point $30\arcsec$ West of this center.  The bottom row corresponds to a point $30\arcsec$ South of this center.  The distributions remain almost identical as the pie slices are moved perpendicular to their orientation.  The area between the opposing radial distributions begins to grow as the pie slices are moved away from the true center of the cluster parallel to their orientation.  The coordinates are in projected distance with respect to the center in the catalog (Harris 1996).}
\label{fig8}
\end{figure}

\clearpage

\begin{deluxetable}{llcrcrr}
\tablewidth{0pc}
\tablecaption{Centers of Milky Way Globular Clusters \label{table1}}
\tablehead{
\colhead{Cluster ID}  & \colhead{Alternate ID} & \colhead{RA J2000}  & \colhead{Dec J2000}  & \colhead{Estimated}  & \colhead{$\ell$} & \colhead{$b$}\\
\colhead{} & \colhead{} & \colhead{( h : m : s )} & \colhead{( $\degr$ : $\arcmin$ : $\arcsec$ )} & \colhead{Uncertainty ($\arcsec$)} & \colhead{($\degr$)} & \colhead{($\degr$)}
 }
 \startdata

NGC 104 &  47 Tuc	&  00:24:05.71  &  -72:04:52.7  &    0.2  &  282.7334  &  -29.5475\\
NGC 288 & 		&  00:52:45.24  &  -26:34:57.4  &    1.8  &  308.4303  &   36.0769\\
NGC 362 & 		&  01:03:14.26  &  -70:50:55.6  &    0.1  &  316.5174  &  -12.8101\\
NGC 1261 & 		&  03:12:16.21  &  -55:12:58.4  &    0.1  &  316.7442  &  -60.6231\\
Palomar 1 & 		&  03:33:20.04  &  +79:34:51.8  &    0.8  &  133.3390  &   22.7027\\
Palomar 2 & 		&  04:46:05.91  &  +31:22:53.4  &    0.1  &  103.5433  &  -28.4620\\
NGC 1851 & 		&  05:14:06.76  &  -40:02:47.6  &    0.1  &  248.3607  &  -20.6108\\
NGC 2298 & 		&  06:48:59.41  &  -36:00:19.1  &    0.1  &  243.1166  &  -22.4883\\
NGC 2808 & 		&  09:12:03.10  &  -64:51:48.6  &    0.1  &  328.9732  &  -19.2632\\
E 3 & 	       		&  09:20:57.07  &  -77:16:54.8  &    1.5  &  316.8235  &  -30.9358\\
NGC 3201 & 		&  10:17:36.82  &  -46:24:44.9  &    1.0  &  278.1893  &    9.2654\\
NGC 4147 & 		&  12:10:06.30  &  +18:32:33.5  &    0.1  &   17.7755  &   65.4396\\
NGC 4590 &  M 68 	&  12:39:27.98  &  -26:44:38.6  &    0.2  &   22.8622  &  -50.2242\\
NGC 4833 & 		&  12:59:33.92  &  -70:52:35.4  &    0.3  &  289.7864  &  -43.0012\\
NGC 5024 &  M 53	&  13:12:55.25  &  +18:10:05.4  &    0.1  &  192.5475  &    0.2276\\
NGC 5053 & 		&  13:16:27.09  &  +17:42:00.9  &    1.4  &  198.6739  &   11.2637\\
NGC 5139 & $\omega$ Cen	&  13:26:47.28  &  -47:28:46.1  &    0.1  &  273.0120  &    4.0733\\
NGC 5272 & M 3		&  13:42:11.62  &  +28:22:38.2  &    0.2  &   43.0822  &   80.7444\\
NGC 5286 & 		&  13:46:26.81  &  -51:22:27.3  &    0.1  &  319.1657  &    8.1490\\
NGC 5466 & 		&  14:05:27.29  &  +28:32:04.0  &    0.6  &   61.6991  &    6.5728\\
NGC 5904 & M 5		&  15:18:33.22  &  +02:04:51.7  &    0.0  &  329.0530  &   62.5516\\
NGC 5927 & 		&  15:28:00.69  &  -50:40:22.9  &    0.2  &  330.6481  &    1.7546\\
NGC 5986 & 		&  15:46:03.00  &  -37:47:11.1  &    0.2  &    4.1513  &  -35.0802\\
Lyng\aa \ 7 & 		&  16:11:03.65  &  -55:19:04.0  &    0.9  &  274.5328  &  -55.0366\\
NGC 6093 & M 80		&  16:17:02.41  &  -22:58:33.9  &    0.2  &  219.5801  &  -43.8833\\
NGC 6101 & 		&  16:25:48.12  &  -72:12:07.9  &    0.5  &  282.8777  &  -27.5859\\
NGC 6121 & M 4		&  16:23:35.22  &  -26:31:32.7  &    0.4  &  231.7876  &  -23.6896\\
NGC 6144 & 		&  16:27:13.86  &  -26:01:24.6  &    0.6  &  236.1344  &  -12.1475\\
NGC 6171 & M 107	&  16:32:31.86  &  -13:03:13.6  &    0.1  &  233.5681  &   10.4371\\
NGC 6205 & M 13		&  16:41:41.24  &  +36:27:35.5  &    0.1  &  186.8649  &   57.7933\\
NGC 6218 & M 12		&  16:47:14.18  &  -01:56:54.7  &    0.8  &  272.9232  &   57.2127\\
NGC 6254 & M 10		&  16:57:09.05  &  -04:06:01.1  &    0.1  &  339.7755  &   52.5219\\
NGC 6304 & 		&  17:14:32.25  &  -29:27:43.3  &    0.2  &    4.9970  &  -10.3158\\
NGC 6341 & M 92		&  17:17:07.39  &  +43:08:09.4  &    0.3  &   74.5945  &   13.8815\\
NGC 6352 & 		&  17:25:29.11  &  -48:25:19.8  &    0.6  &  350.4569  &  -44.6173\\
NGC 6362 & 		&  17:31:54.99  &  -67:02:54.0  &    0.5  &  318.4562  &  -46.5374\\
NGC 6366 & 		&  17:27:44.24  &  -05:04:47.5  &    1.4  &   52.7583  &  -42.8224\\
NGC 6388 & 		&  17:36:17.23  &  -44:44:07.8  &    0.3  &  328.0561  &  -70.0233\\
NGC 6397 & 		&  17:40:42.09  &  -53:40:27.6  &    0.3  &  296.6536  &  -63.2206\\
NGC 6441 & 		&  17:50:13.06  &  -37:03:05.2  &    0.2  &  239.5334  &  -54.6078\\
NGC 6496 & 		&  17:59:03.68  &  -44:15:57.4  &    1.3  &  250.5333  &  -29.8676\\
NGC 6535 & 		&  18:03:50.51  &  -00:17:51.5  &    0.4  &  213.9027  &    1.2416\\
NGC 6541 & 		&  18:08:02.36  &  -43:42:53.6  &    0.1  &  258.8651  &   -7.1202\\
NGC 6584 & 		&  18:18:37.60  &  -52:12:56.8  &    0.2  &  283.3375  &    5.5601\\
NGC 6624 & 		&  18:23:40.51  &  -30:21:39.7  &    0.1  &  288.7654  &   30.9569\\
NGC 6637 & M 69		&  18:31:23.10  &  -32:20:53.1  &    0.1  &  317.2313  &   28.8707\\
NGC 6652 & 		&  18:35:45.63  &  -32:59:26.6  &    0.1  &  331.1239  &   22.9512\\
NGC 6656 & M 22		&  18:36:23.94  &  -23:54:17.1  &    0.8  &  338.5181  &   29.5142\\
NGC 6681 & M 70	        &  18:43:12.76  &  -32:17:31.6  &    0.1  &  350.1062  &    8.1499\\
NGC 6715 & M 54		&  18:55:03.33  &  -30:28:47.5  &    0.1  &    9.7305  &  -24.3068\\
NGC 6717 & Pal 9	&  18:55:06.04  &  -22:42:05.3  &    0.2  &   17.7368  &  -21.8003\\
NGC 6723 & 		&  18:59:33.15  &  -36:37:56.1  &    0.3  &    6.3887  &  -39.3572\\
NGC 6752 & 		&  19:10:52.11  &  -59:59:04.4  &    0.1  &  317.8388  &  -55.1195\\
NGC 6779 & M 56		&  19:16:35.57  &  +30:11:00.5  &    0.2  &  127.4063  &  -32.5365\\
Terzan 7 & 		&  19:17:43.92  &  -34:39:27.8  &    0.3  &  260.5393  &  -79.4448\\
Arp 2 & 		&  19:28:44.11  &  -30:21:20.3  &    0.9  &  229.4959  &  -46.4803\\
NGC 6809 & M 55		&  19:39:59.71  &  -30:57:53.1  &    0.8  &  241.7989  &  -11.9084\\
Terzan 8 & 		&  19:41:44.41  &  -33:59:58.1  &    1.4  &  247.0191  &   -8.2681\\
NGC 6838 & M 71		&  19:53:46.49  &  +18:46:45.1  &    0.5  &  219.2432  &   55.7621\\
NGC 6934 & 		&  20:34:11.37  &  +07:24:16.1  &    0.1  &   51.9165  &  -18.6097\\
NGC 6981 & M 72		&  20:53:27.70  &  -12:32:14.3  &    0.1  &  150.5005  &  -73.7548\\
NGC 7078 & M 15		&  21:29:58.33  &  +12:10:01.2  &    0.2  &  230.1238  &   53.5815\\
NGC 7089 & M 2		&  21:33:27.02  &  -00:49:23.7  &    0.1  &  261.6423  &   54.6937\\
NGC 7099 & M 30		&  21:40:22.12  &  -23:10:47.5  &    0.1  &  307.1349  &   39.5768\\
Palomar 12 & 		&  21:46:38.84  &  -21:15:09.4  &    0.4  &  334.1046  &   34.9822\\

\enddata
 \end{deluxetable}

   \end{document}